\documentstyle[aps,epsfig,multicol,fancyheadings]{revtex}
\begin{document}

\title{Intermittence and roughening of periodic elastic media}

\author{E.\ T.\ Sepp\"al\"a,$^1$  M.\ J.\ Alava,$^1$ and P.\ M.\ Duxbury$^2$}

\address{$^1$Helsinki University of Technology, Laboratory of Physics, 
P.O.Box 1100, FIN-02015 HUT, Finland}
\address{ $^2$Dept. of Physics and Astronomy and Center for 
Fundamental Materials Research,\\
Michigan State Univ., E. Lansing, MI 48824-1116}
\date{\today}

\maketitle

\begin{abstract}
We analyze intermittence and roughening of an elastic interface or
domain wall pinned in a periodic potential, in the presence of
random-bond disorder in (1+1) and (2+1) dimensions.  Though the
ensemble average behavior is smooth, the typical behavior of a large
sample is intermittent, and does not self-average to a smooth
behavior.  Instead, large fluctuations occur in the mean location of
the interface and the onset of interface roughening is via an
extensive fluctuation which leads to a jump in the roughness of order
$\lambda$, the period of the potential. Analytical arguments based on
extreme statistics are given for the number of the minima of the
periodicity visited by the interface and for the roughening
cross-over, which is confirmed by extensive exact ground state
calculations.
\end{abstract}

\noindent {\it PACS \# \ \ 05.70.Np, 75.10.Nr, 02.60.Pn, 68.35.Ct}
\begin{multicols}{2}[]
\narrowtext

\section{Introduction}\label{intro}

The properties of extended, elastic manifolds like domain walls in
magnets or contact lines of liquids on solid substrates becomes very
varied if one introduces some disorder. Defects on a surface or
impurities in a magnet often pin such interfaces. The recent interest
in their physics follows from the observation that the energetics in
the presence of randomness is obtained by optimizing the configuration
of the manifold \cite{HaH95}. A competition between elasticity and the
random potential arises. It results in scale-invariance described by
the {\it roughness exponent} that measures the geometrical
fluctuations, and the {\it energy fluctuation exponent} that measures
the variation of the manifold energy around its mean. It is also
related to the energy scales of excitations from the state of minimum
energy.  The experimental interest in these systems arises in
particular due to the energetics: time-dependent phenomena like creep
and coarsening (in magnets) follow slow, activated dynamics dictated
by the energy barriers that can be described with such exponents
\cite{Lemerle98}.

Frequently manifolds also experience a periodic potential.  In the
case of superconductors, one periodicity is due to the rotational
invariance of the phase.  A second periodicity is induced when flux
lines form a lattice. Similarly, in the case of charge density waves
(CDW) or domain walls in magnets, one periodicity is due to the
underlying lattice structure, and a second is due to the
self-organized periodicity of the CDW or magnetic domains themselves.
Generic models for these phenomena are called {\it periodic elastic
media} (PEM) and are the focus of this work.  As noted recently the
asymptotic behavior of the PEM class depends on the type of
periodicity, with the case of a periodic surface tension being in one
universality class\cite{Eira2000}, while the case of an applied
periodic potential is in another\cite{Bouchaud92}.  In this work we
are interested in the case of an applied periodic potential, in
particular the intermittent behavior of interfaces which experience a
competition between pinning due to the periodic potential and pinning
due to random bond disorder.
 
The paper is organized as follows. Sec.~\ref{PEM} introduces the
Hamiltonian of periodic elastic media and describes two intermittent
behaviors involved in PEM, when the amplitude of the applied
periodicity is changed. The Section also includes a discussion of the
numerical method used.  In Sec.~\ref{interm1} the first type of the
intermittent behavior, jumps of manifolds, is analyzed using extremal
statistics and is demonstrated with numerical simulations.
Sec.~\ref{interm2} discusses the second type of the intermittent
behavior, the roughening of the manifolds, with the aid of droplet
arguments and further numerics.  In Sec.~\ref{100} the roughening
behavior is studied in $\{10\}$ and $\{100\}$ oriented lattices which
have a lattice induced periodicity; we compare these systems with
other PEM. The paper is finished with conclusions in Sec.~\ref{concl}.

\section{Periodic Elastic Media}\label{PEM}

The continuum Hamiltonian that describes the competition between
elasticity, periodicity and randomness is given by,
\begin{equation}
{\mathcal H}_{pem} = \int \left[ {\gamma\over 2}  \{\nabla h(\vec{r})\}^2 
+ \eta\{h(\vec{r})\} + V_p\{h(\vec{r})\} \right] d\vec{r}.
\label{eq_pem} 
\end{equation}
Here $h(\vec{r})$ is a single valued height variable, and $\vec{r}$ is
a $(d-1)$-dimensional vector.  $V_p$ is a periodic potential (of
amplitude $V_0$ and wavelength $\lambda$) in the height direction and
$\eta\{h(\vec{r})\}$ is the disorder, which we take to be of the
random bond type with delta-function correlations. The physics of
manifolds described by Eq.~(\ref{eq_pem}) has been discussed recently
since there may exist a {\it roughening transition} that separates an
algebraically rough regime from a logarithmically rough one as the
potential strength is varied \cite{Bouchaud92}.  However, in the
dimensions considered here [$d = (1+1), (2+1)$] these manifolds are
always rough at large enough length scales \cite{Fisher86}.  The
issues we raise here arise in all dimensions, and so we numerically
illustrate them in ($1+1$)- and ($2+1$)-dimensional systems.

We calculate the exact location and morphology of interface
ground-states for a given configuration of bond disorder.  For this
configuration of bond disorder we vary the {\it amplitude} of the
periodic modulation.  Interfaces which experience this combination of
a periodic potential and random bond disorder show a variety of
intermittent behavior as the amplitude of the potential, $V_0$, is
varied.  Two types of intermittence which we study in detail are:
intermittent jumps in the center of mass location of the interface;
and intermittent jumps in the roughness of the interface.  The first
type is most easily discussed at strong pinning (large values of the
key ratio $v = V_0 \lambda J/\delta J$) where the interface is always
pinned near a minimum of the periodic potential, but it jumps between
different minima as $v$ is varied.  It does this to maximize the
energy gain due to small fluctuations about a flat interface.  In the
limit of large system sizes there can be an infinite number of such
jumps with, of course, no overlap between the ground states of
interfaces in different minima.  We develop a scaling theory to
demonstrate that the number of minima explored as $v$ is varied over a
finite range is of order $\ln(L_h)$, where $L_h$ is the system size in
the $h$-direction in which the manifold fluctuates.  Such
intermittence is similar to the chaos seen in spin glasses (where it
implies vanishing overlap between spin configurations) and is related
to replica symmetry breaking \cite{Parisi90}.  It is also a close
cousin of the phenomenon that takes place if the disorder is changed
randomly \cite{Zhang87}.

A second type of intermittence occurs when it becomes energetically
favorable to form a large domain excitation.  This means that a finite
fraction of the interface is in one minimum of the potential, while
another finite fraction is in an adjacent minimum.  These large
fluctuations are the classical ``Imry-Ma'' -type droplets and have a
linear extension of the order of the system size.  We are, by slowly
decreasing the potential, able to find the threshold at which the
first domain excitation occurs and to demonstrate its effect on the
roughness $w(v)$.  We observe that since the domain excitation is of
the order of the sample size, the roughness produced by that domain
fluctuation is proportional to $\lambda$.  Thus there is a {\it
first-order} jump in roughness.  In contrast, a naive averaging of the
roughness looks smooth and scales nicely.  This is due to a scaling of
the {\it probability} of a jump of the order $\lambda$ occurring at
$v$ rather than being the self-averaging behavior of a typical
sample. The exact numerical calculations are supported by scaling
theories based on Imry-Ma and large fluctuation ideas which account
for the jumpy behavior of interfaces in a periodic potential.

The numerical calculations are carried out using Ising magnets with
random bonds.  For a given configuration of bond disorder, we find the
ground state interface in square and cubic, nearest-neighbor,
spin-half, ferromagnetic Ising models.  An interface is imposed along
the $\{11\}$ or $\{10\}$ directions of a square lattice or along the
$\{111\}$ or $\{100\}$ directions of a cubic lattice, by using
antiperiodic boundary conditions.  Periodic boundaries are used
directions parallel to the interface unless otherwise mentioned.  The
average value of the exchange constant is $J =1$, while the
random-bond disorder is drawn from a uniform distribution of width
$\delta J$.  The periodic potential $V_p = V_0 [0.5\sin (2\pi
h/\lambda) +0.5]$ is added to the random bond disorder, where $h$ is
to be along a direction perpendicular to the average orientation of
the interface. This is done for the $\{11\}$ and $\{111\}$ cases while
in the other orientations ($\{10\}$, $\{100\}$) we use the intrinsic
lattice potential as discussed below.  Notice that if $\lambda$ is
small, the discrete representation of the potential will by necessity
be rather coarse. The exact interface ground state in this random
energy landscape is found using a mapping to the minimum-cut
maximum-flow optimization problem \cite{Alavaetal}.  We have developed
a highly efficient (in both memory and speed) implementation of the
push-and-relabel method for the maximum flow problem
\cite{Goltar88}. The exact ground state of a manifold in system with
one million sites can be found in about a minute on a workstation.
 
\section{Jumps Between Potential Valleys}\label{interm1}

We first discuss the sensitivity of the ground state of the model
(\ref{eq_pem}) to small variations in the amplitude $V_0$ of the
potential $V_p$ with wave length $\lambda$.  A simple scaling theory
captures many aspects of this sensitivity.  The scaling theory begins
with the central limit form for the energy of a flat interface located
at a minimum of the periodic potential, $P_1(E)$.  If the interface is
exactly flat, the energy fluctuations are just due to the random bond
disorder, so that
\begin{equation}
P_1(E) =  \frac{1}{\sqrt{\pi} \sigma} \exp \left \{ -{(E - J A)^2\over  
\sigma^2} \right \},
\label{distr}
\end{equation}
where $A=L^{d-1}$ is the area of the manifold and $\sigma^2 = 2 A
\delta J^2$ is the width of Gaussian distribution.

Consider now a system in which there are $N$ minima in the periodic
potential.  The probability, $L_N(E)$, that the lowest minima has
energy $E$ is, $L_N(E) = N P_1(E) \{1-C_1(E)\}^{N-1}$ where, $C_1(E) =
\int_{-\infty}^E P_1(e)\, de.$ The difference in energy, $g$, between
the lowest energy state and the next lowest energy state of the
manifold may also be simply calculated. We call this difference in
energy the ``gap'' and its distribution, $G_N(g,E)$ is given by,
$G_N(g,E) = {N(N-1)} P_1(E) P_1(E+g)\{1-C_1(E+g)\}^{N-2}$.  Stated
more precisely $G_N(g,E)$ is the probability that if the lowest energy
manifold has energy $E$, then the gap to the next lowest energy level
is $g$.  The average lowest energy level is given by $\langle E_M
\rangle = \int_{-\infty}^{\infty} E L_N(E) \, dE$.  This is not
analytically tractable.  However, the typical value of this lowest
energy is estimated from $\sigma N P_1(\langle E_M \rangle )\approx 1$
which yields,
\begin{equation}
\langle E_M \rangle \sim JA - \sigma \{ \ln(N)\}^{1/2} 
\label{typicalene}
\end{equation}

To estimate the typical value of the gap, we use, $\sigma^2 {N(N-1)}
P_1(\langle E_M \rangle ) P_1(\langle E_M \rangle + \langle g \rangle)
\approx 1$, which with (\ref{typicalene}) and the fact that $| \langle
g \rangle | \ll | \langle E_M \rangle |$ yields,
\begin{equation}
\langle g \rangle \approx {\sigma^2 \ln(\sigma) \over (JA
- \langle E_M \rangle )} \approx {\sigma \ln(\sigma) \over \{ \ln(N)\}^{1/2}},
\label{gap}
\end{equation}
where $\sigma = \sqrt{2 A} \delta J$ and $A=L^{d-1}$.  The gap between
minima of the potential is thus of order $1/\{\ln(N)\}^{1/2}$, where
$N \sim L_h/\lambda$ and $L_h$ is the system size perpendicular to the
interface.  So the separation between minima grows increasingly small
as $L_h$ increases. Similar extreme statistics problems are discussed
in \cite{Galambos}.

Given the small gap between the metastable minima of the periodic
potential, due to the presence of random bonds, we now need to find
the typical change in $V_0$ which can cause a level crossing in which
the global ground state changes from one minima of the periodic
potential to another \cite{Seppala00}.  The key new effect that we
must control is the fact that the interfaces are {\it not flat} even
when confined to one minimum of the periodic potential.  Instead they
have a roughness which is determined by the interplay between the
curvature of the periodic potential at its minima and the energy
variations of a manifold due to confinement.  We now develop a scaling
theory for this phenomenon.

First we treat the confinement effect.  Consider a manifold in the
presence of random bond disorder and which is confined in a slab of
size $l \times L^{d-1}$.  The energy of such a slab is given by,
\begin{equation}
E(l,L) = \left({L\over L_x}\right)^{d-1} \left( c_1 L^{d-1}_x 
+ c_2 L_x^{\theta}\right),
\label{ene_l}
\end{equation}
where $L_x = l^{1/{\zeta}}$.  This yields,
\begin{equation}
\epsilon(l) = {E(l,L)\over L^{d-1}} = c_1 + c_2 l^{-x},
\label{epsilon}
\end{equation}
where,
\begin{equation}
x=(d-1-\theta)/\zeta
\label{x_eskp}
\end{equation}
Notice that $x$ is {\it positive} so that the confinement energy {\it
decreases} as the confinement length $l$ increases, as expected.

To include the effect of the confining potential, consider the
behavior near a minima of the periodic potential to be of the form,
\begin{equation}
V(l) = {\mathcal V}_0 \left( {l\over \lambda}\right)^y,
\label{v_l}
\end{equation}
where ${\mathcal V}_0 = V_0/\delta J$, and $y$ is a positive exponent
to ensure that the potential is confining. For example a sinusoidal
potential has $y=2$.  The behavior of a manifold in this confining
potential and in the presence of an additive random bond disorder, is
estimated by considering its total energy as a function of $l$
[i.e. combining (\ref{ene_l}) and (\ref{v_l})],
\begin{equation}
\epsilon_{total} = c_1 +c_2l^{-x} + {\mathcal V}_0 \left({l\over 
\lambda}\right)^y,
\label{epsilon_tot}
\end{equation}
Finding the minimum of the total energy yields the manifold roughness,
\begin{equation}
l_c = \left({c_2 x \lambda^y \over y {\mathcal V}_0} \right)^{1/(y+x)},
\label{l_c}
\end{equation}
with the energy of this optimal manifold being
\begin{equation}
\epsilon_{opt} = c_1 + c_3 \left({c_2^y  {\mathcal V}_0^x 
\over \lambda^{xy}} \right)^{1/(y+x)},
\label{epsilon_opt}
\end{equation}
where $c_3$ is a constant that depends on $x$ and $y$ \cite{fisu}.

Now the variation in the optimal energy with a small variation in
${\mathcal V}_0$ is given by,
\begin{equation}
\epsilon_{opt} ({\mathcal V}_0 +\delta {\mathcal V}_0) -  \epsilon_{opt} 
({\mathcal V}_0 ) = {\partial \epsilon_{opt}\over \partial {\mathcal V}_0} 
\delta {\mathcal V}_0
\label{epsilon_var}
\end{equation}
This {\it change in energy} also varies randomly from one minimum of
the potential to another.  If {\it the variation in the energy change}
is of order the gap found in Eq.~(\ref{gap}), then we expect the
ground state location to change from one minimum of the potential to
another.  Thus we find the typical value of $\delta {\mathcal V}_0$
between jumps to be found from,
\begin{equation}
\left (L^{d-1} {\partial \epsilon_{opt}\over \partial {\mathcal V}_0} \delta
 {\mathcal V}_0 \right)^{1/2} = \langle g \rangle.
\end{equation}
Thus using Eq.~(\ref{gap}),
\begin{eqnarray}
\delta V_{0jump} = {\langle g \rangle^2\over L^{d-1}} \left({\partial 
\epsilon \over \partial {\mathcal V}_0} \right)^{-1} = {\langle g 
\rangle^2\over L^{d-1}} {(x+y)\over c_3 x} \left({\lambda^x {\mathcal V}_0
\over c_2} \right)^{y/(x+y)} 
\nonumber \\
\sim {\delta J^2 \left\{ \ln \left(L^{d-1}\delta J^2 \right) \right\}^2
\over \ln(N)} \left({\lambda^x V_0\over \delta J} \right)^{y/(x+y)},
\label{dV0_jump}
\end{eqnarray}
where ${\mathcal V}_0 = V_0/\delta J$.  There are several interesting
features of this equation.  Firstly note that $\delta V_{0jump}$
increases logarithmically with area of the manifold $L^{d-1}$. On the
other hand, the number of minima $N \sim L_h/\lambda$, and $\delta
V_{0jump}$ decreases logarithmically with $L_h$.  The dependence of
$\delta V_{0jump}$ on $\lambda$ and on $V_0$ is qualitatively as
expected in that it increases monotonically with both of them.

\begin{figure}  
\centerline{\epsfig{file=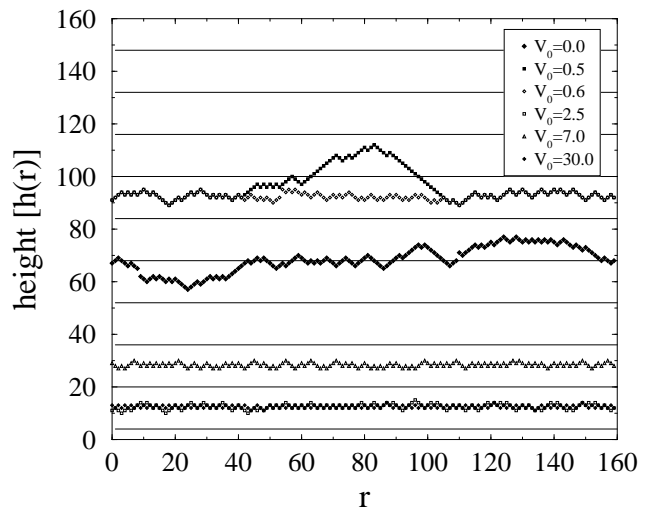,width=7cm,angle=-90}}
\caption{ Interface configurations in $(1+1)$-dimensions for various
$V_0$.  In this calculation the disorder configuration and wavelength
($\lambda = 16$) are fixed at $\delta J = 1$.  As $V_0$ is varied, the
interface jumps between the minima of the periodic potential.  The
solid lines denote the position of the largest values of the
sinusoidal periodic potential $V_p$. The lattice size is $160 \times
160$ and the interfaces are oriented along the $\{11\}$ direction.
Note that the disorder is {\it exactly} the same for each value of
$V_0$.}
\label{fig1}
\end{figure}

\begin{figure}
\centerline{\epsfig{file=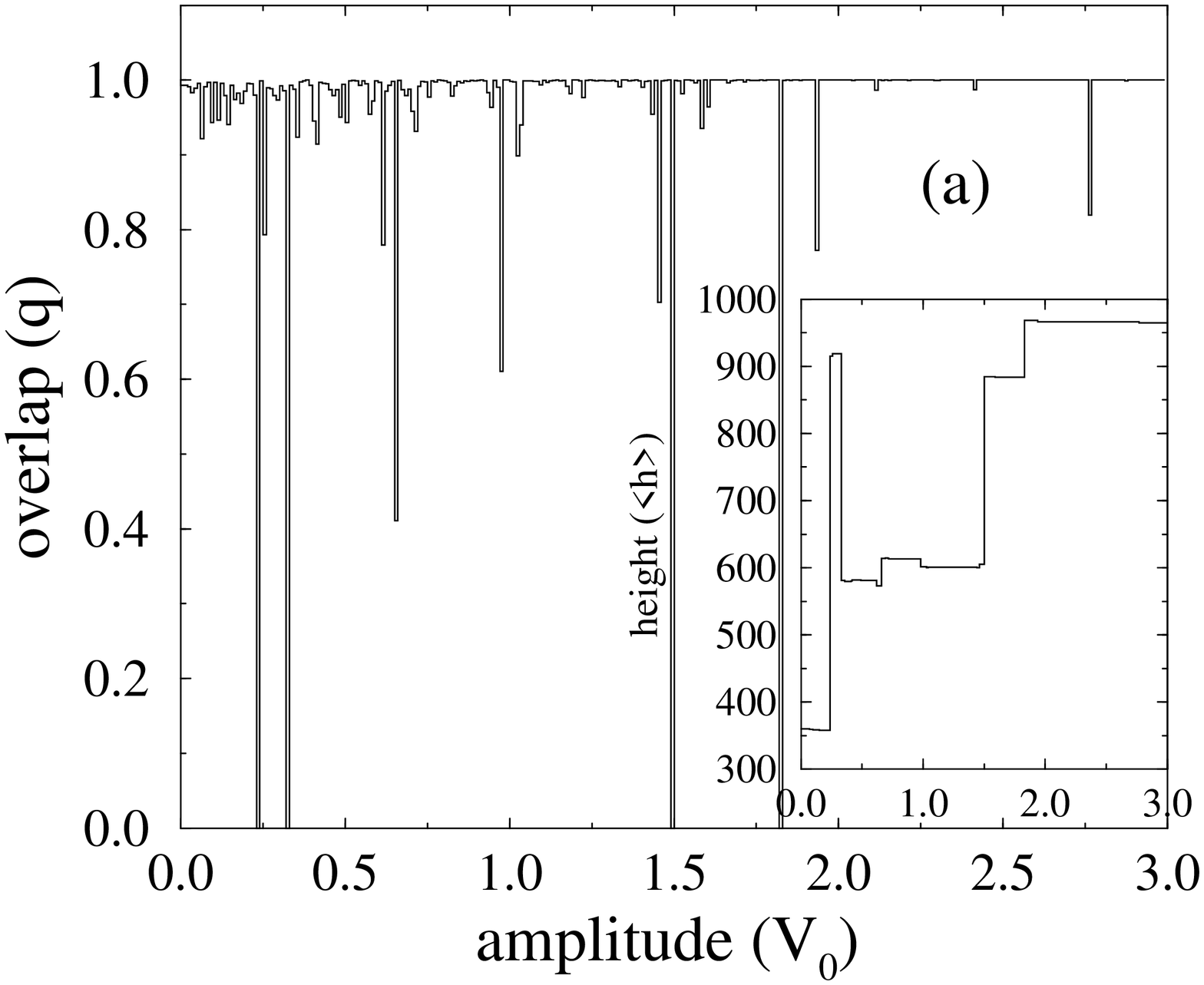,width=7cm,angle=-90}}
\centerline{\epsfig{file=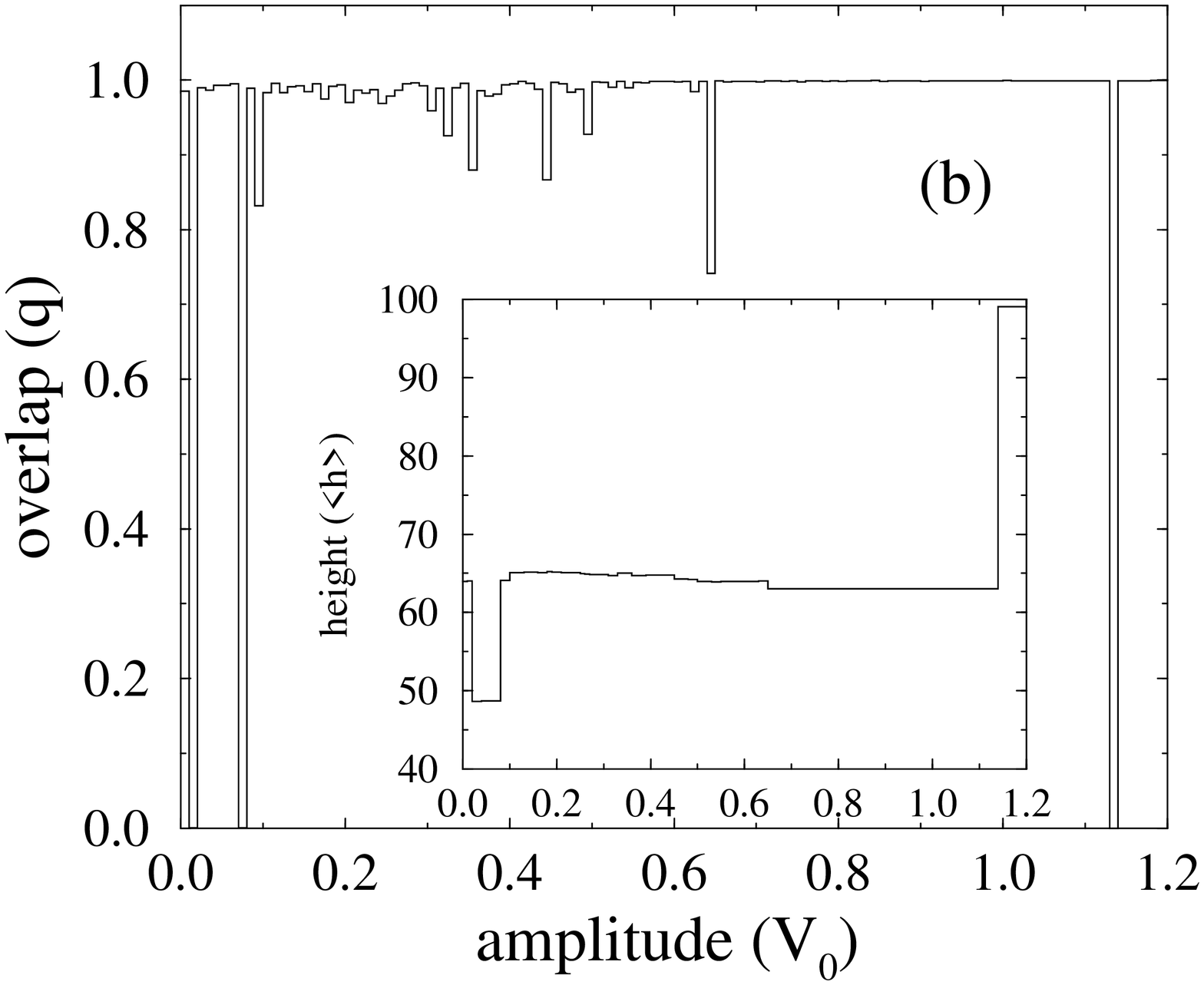,width=7cm,angle=-90}}
\caption{The overlap $q = L^{-(d-1)}\sum_i \delta(h_i^1 - h_i^2)$
between ground states as the amplitude of the potential $V_0$ is
varied ($\delta J =1$). As $V_0$ is decreased, we calculate the
overlap between the interface configuration at one value of $V_0$
(described by \{$h_i^1$\}) and the interface at the next value of
$V_0$ (described by \{$h_i^2$\}). The corresponding mean heights
$\langle h \rangle$ are shown in the insets. The calculations were
carried out as for Fig.~\ref{fig1}, however we used 300 different
values of $V_0$ with $\Delta V_0 =10^{-2}$ for the same realisation of
disorder and the wavelength $\lambda=4$.  (a) Two dimensional case,
with the system size $L \times L_h = 1024 \times 1025$.  (b)
$(2+1)$-dimensional interfaces oriented along the $\{111\}$ direction
for lattices of size $L^2 \times L_h = 100^2 \times 129$.}
\label{fig2}
\end{figure}

The intermittence implied by the result (\ref{dV0_jump}) is
illustrated in Fig.~\ref{fig1} and in Fig.~\ref{fig2}.  As a function
of $V_0$, the manifold mostly stays almost unchanged in the current
valley of minimum energy and occasionally jumps to another, new
minimum of the periodic potential.  A useful way to illustrate this
intermittence as a function of $V_0$ is to calculate the {\it
configurational overlap} between the ground states as a function of
$V_0$ (in analogy with the overlap used in spin-glasses \cite{Young}).
The overlap, $q$ is one if two configurations are the same and zero if
they have no bonds in common.  Fig.~\ref{fig2} presents the overlap as
a function of amplitude of the pinning potential, $V_0$, for
interfaces in square and cubic lattices.  The intermittent nature of
periodic elastic media is clearly evident in these figures.  Note that
while the overlap and the interface roughness are intermittent, the
{\it interface energy} (see Fig.~\ref{fig3}) does not show any obvious
signs of the jumps.  Due to the logarithmic reduction in the gap size
[Eq.~(\ref{gap})], the interface will only sample an infinitesimal
fraction [$\ln(N)/\{\ln(L^{d-1})\}^2$] of the available minima of the
potential as we sweep $v$.  Nevertheless a large number of different
minima [$\sim \ln(L_h)$] will be sampled by the system in particular
if $L_h$ is increased while the transverse size $L$ is kept fixed.

\section{Roughening of the Manifolds}\label{interm2}

\begin{figure}
\centerline{\epsfig{file=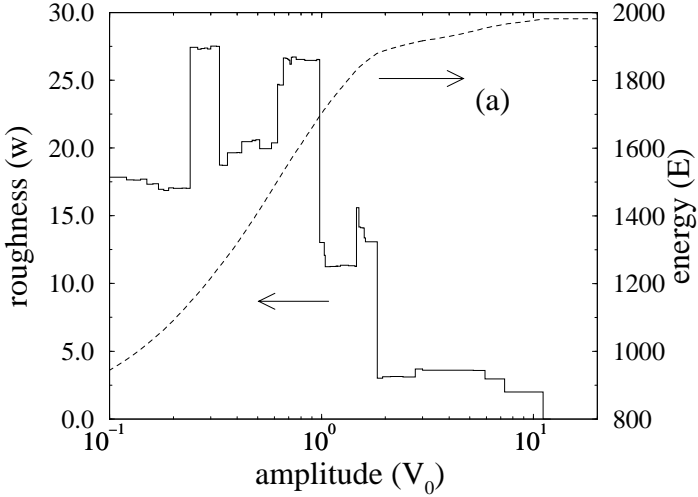,width=7cm,angle=-90}}
\centerline{\epsfig{file=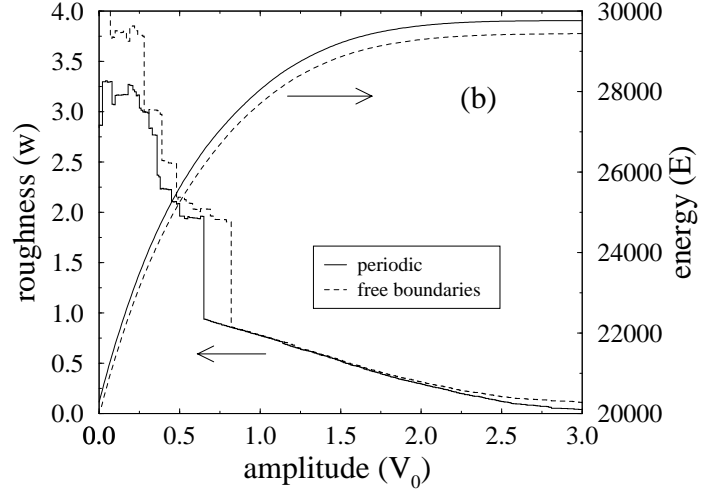,width=7cm,angle=-90}}
\caption{The interface width [$w^2 = L^{-(d-1)} \sum_i(h_i - \langle h
\rangle) ^2$] and the total energy as a function of $V_0$ for $\lambda
= 4$ and $\delta J = 1$.  The results are for a fixed disorder
configuration and from the same calculations as Fig.~\ref{fig2}. (a)
$(1+1)$-dimensional system. (b) $(2+1)$-dimensional system.
The systems with free and periodic boundaries have the same
realization of randomness.}
\label{fig3}
\end{figure}

The behavior of the roughness of interfaces seen in Fig.~\ref{fig3} is
also strongly intermittent, especially in ($1+1$)- dimensions.  The
large jumps in roughness seen in this figure are easily understood
from the Imry-Ma arguments~\cite{Imr75} concerning the instability of
interfaces to large fluctuations, as we now demonstrate for the
($2+1$)-dimensional case.

The interface energy of a subregion $a$ of the interface is, of
course, also drawn from the Gaussian, $P_1(E) = \frac{1}{\sqrt{\pi}
\sigma} \exp \left \{-{(E - J a)^2 \over \sigma^2} \right \}$, but now
with standard deviation $\sigma^2 = 2 a \delta J^2$.  Some of these
energy fluctuations are favorable while others are unfavorable.  The
largest favorable fluctuations are found by setting $a \sigma
P_1(E)\approx 1$, similarly to the extreme statistics arguments as in
deriving Eq.~(\ref{typicalene}), and it gives as the value of the
energy gain,
\begin{equation}
\langle E_g \rangle \approx \sigma  \{\ln(a)\}^{1/2}. 
\label{gain}
\end{equation}

A flat interface would like to take advantage of such large favorable
energy fluctuations in adjacent minima of the periodic potential.
However this requires having segments of the interface crossing the
barriers in the periodic potential.  We define the barrier cost per
bond to be $\epsilon^0_b$ and this is given by the integral over the
barrier, $\epsilon^0_b = {1\over \lambda} \left \{ \int_0^{\lambda}
V(x)\, dx\right \} = \epsilon V_0$.  We shall use the last of these
forms as we shall often be interested in the dependence on $V_0$.  We
consider $(1+1)$ and $(2+1)$ dimensional systems of wavelength
$\lambda$, length $L$ and width $B$, so that $a_b \simeq \lambda B$ is
the area of the part of the interface, which crosses the energy
barrier, and $a \simeq LB/2$ in order to maximize the energy gain.
$B=1$ is the two dimensional case, and $B=L$ in the isotropic three
dimensional case. The barrier energy cost is given by
\begin{equation}
E_b = \epsilon V_0 \lambda B.
\label{cost}
\end{equation}

Equating Eqs. (\ref{gain}) and (\ref{cost}), yields the estimate of the
parameter values at which the first ``Imry-Ma'' jump in the manifold
roughness occurs,
\begin{equation}
\left ({\epsilon V_0 \lambda \over \delta J} \right )_1 
\sim \left [ \sqrt{L \over B} \{ \ln (B L) \}^{1/2} \right ]_1.
\label{Imryjump}
\end{equation}
In the ($1+1$)-dimensional case the logarithmic correction drops
out, by elementary considerations.

An ``Imry-Ma'' fluctuation of size $a$ leads to a jump in the
roughness, which is of order $\lambda \times a/A \simeq \lambda/2$,
since $A=BL$.  We emphasize that this is the expected outcome in any
system with fixed disorder, when $V_0$ is varied.  If $B \propto L$,
there is an exponential dependence of the crossover length on the
parameters, for example, for $B=L$,
\begin{equation}
L_1 \sim  \exp \left [ \left ({\epsilon V_0  \lambda \over \delta J} 
\right)^2 \right ],
\label{crossover_l}
\end{equation}
an exponential dependence on $v$~\cite{Imr75}.

\begin{figure}
\centerline{\epsfig{file=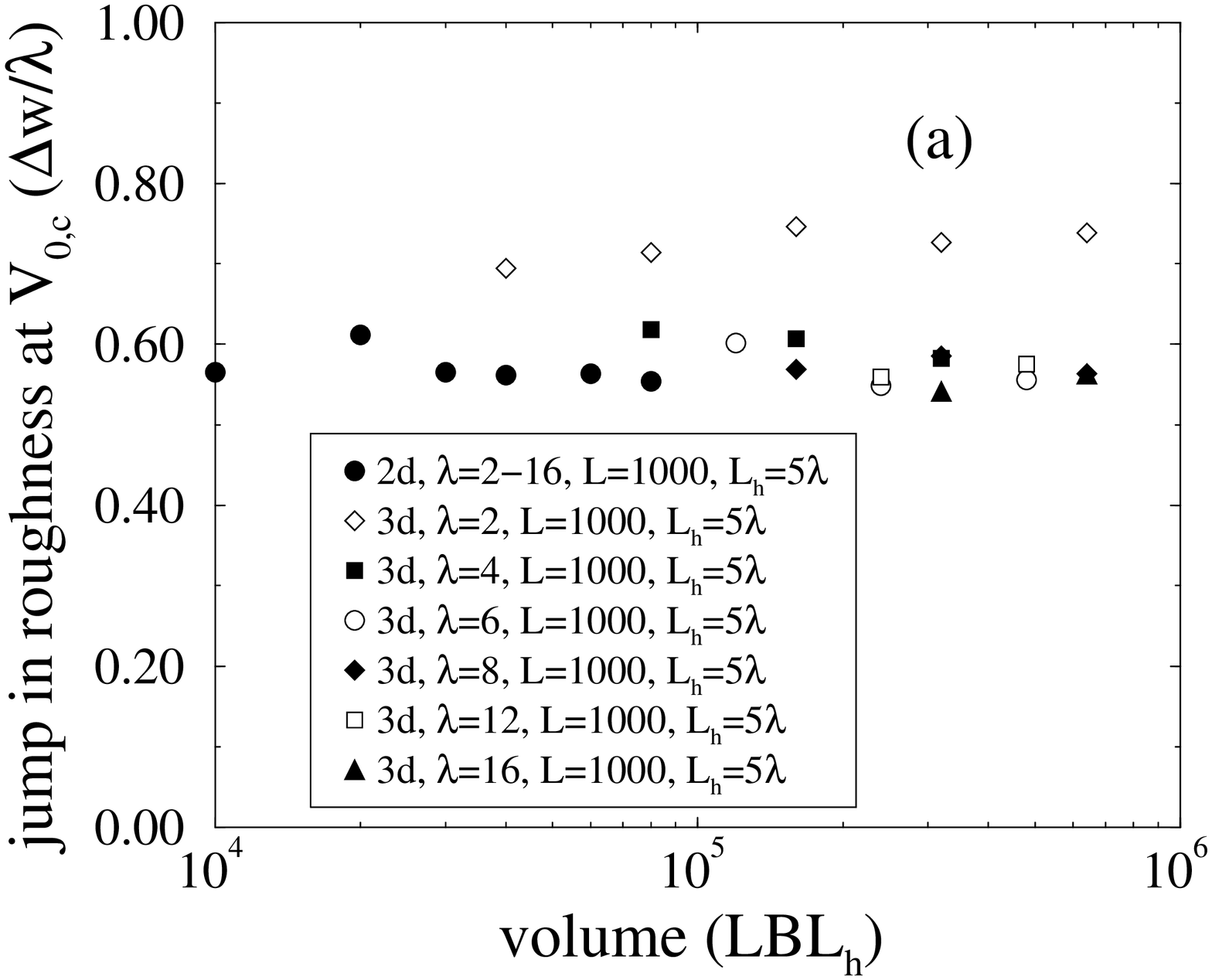,width=7cm,angle=-90}}
\centerline{\epsfig{file=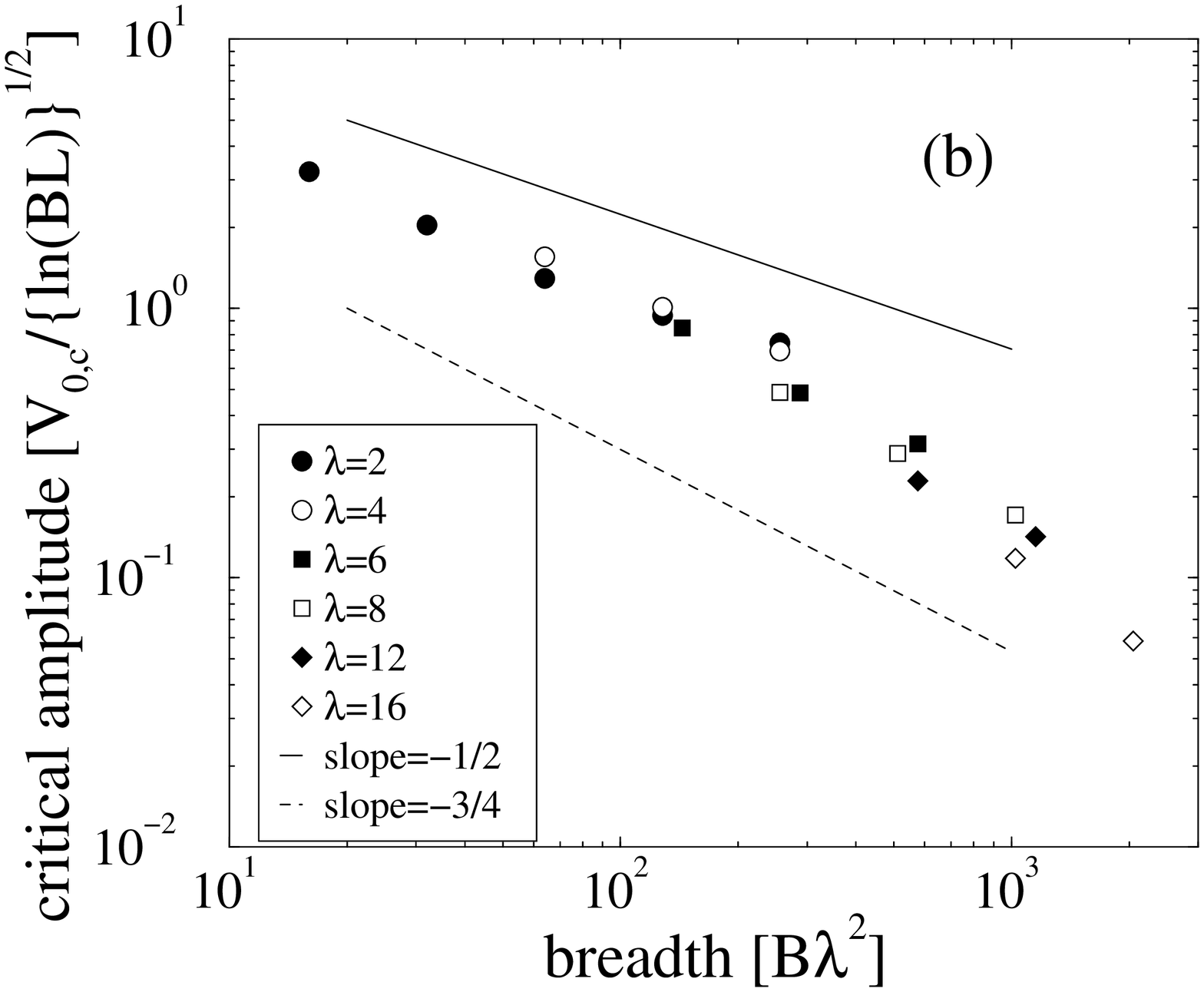,width=7cm,angle=-90}}
\caption{(a) Average size of the first jump in roughness $\Delta w$,
when $V_0=V_{0,c}$, normalized using $\lambda$ and calculated as the
difference between roughness values just after a jump and before that,
as a function of the volume of the systems.  We have carried out
simulations for a strips of dimension $L=1000$, $B=1 \-- 64$, and
$L_h=5 \lambda$, for various values of $\lambda$. The number of
realizations is 100.  (b) Average value of the amplitude of the
potential $V_0=V_{0,c}$ at which the large-scale ``Imry-Ma''
fluctuation occurs ($\delta J =1$).  The data is from the same
simulations as in (a) for $(2+1)$ dimensional case, i.e. $B>1$. The
results are scaled using the prediction (\ref{Imryjump}).}
\label{fig4}
\end{figure}

In Fig.~\ref{fig3}, we present the numerically observed behavior of
the interface roughness as a function of $V_0$. We observe that for
very large $V_0$ the interfaces are flat and are confined to a minimum
of the potential.  For a large range of $V_0$ the roughness stays the
same or increases slowly (in three dimensions), until finally at a
critical value a discrete jump occurs due to the ``Imry-Ma''
nucleation process. This implies that the roughening process, as
defined by the point at which the interfaces begin to fluctuate
outside a single valley, has a {\it first-order character}.  It is
seen from Fig.~\ref{fig4}(a) that the first jump is $\propto \lambda$
as expected for an extensive fluctuation.  The critical value $V_{0,c}$
at which the first extensive fluctuation occurs [Fig.~\ref{fig4}(b)]
follows roughly the prediction of Eq.~(\ref{Imryjump}) though the
slope is closer to 3/4 instead of 2/3. 

\begin{figure}
\centerline{\epsfig{file=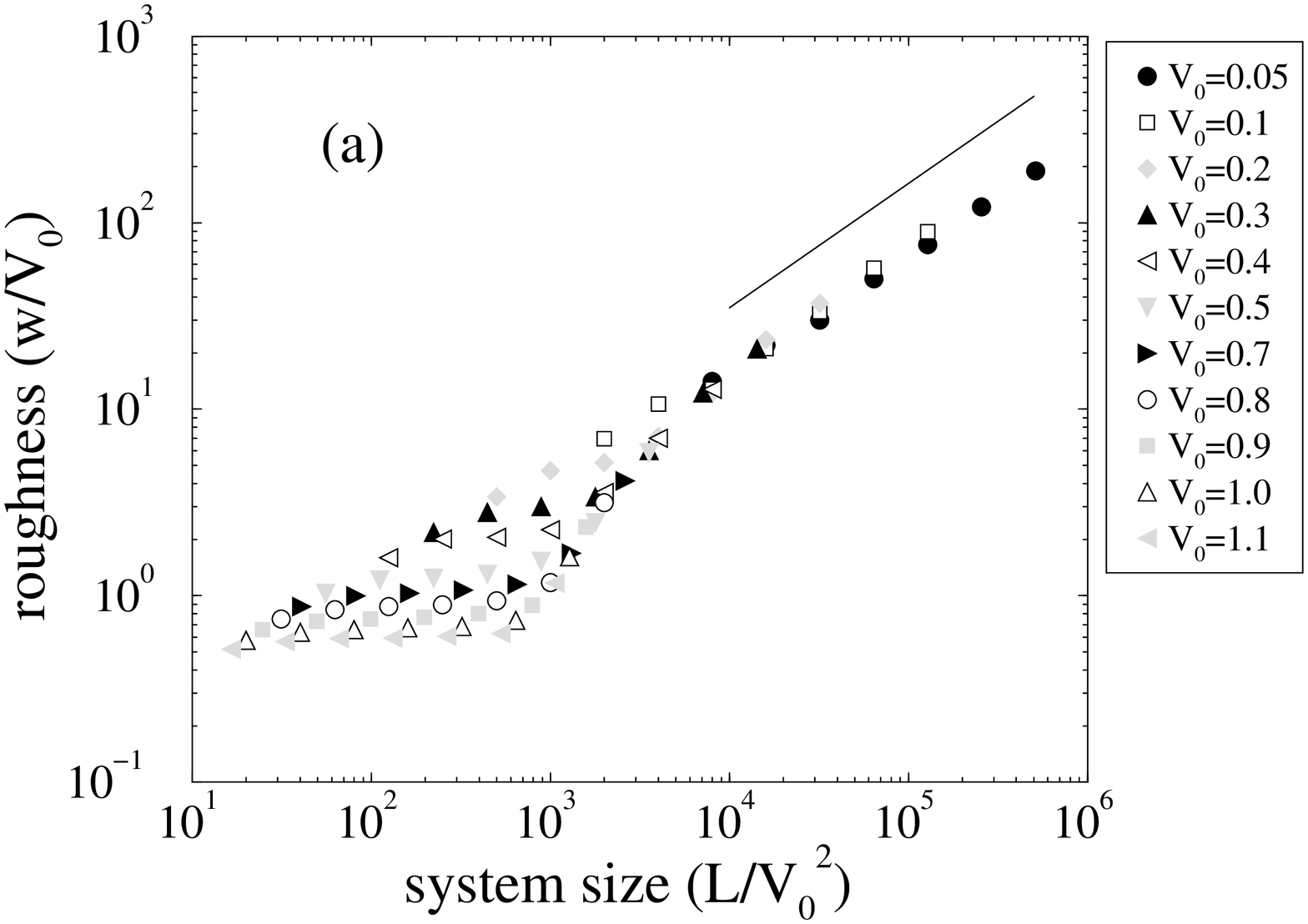,width=7cm,angle=-90}}
\centerline{\epsfig{file=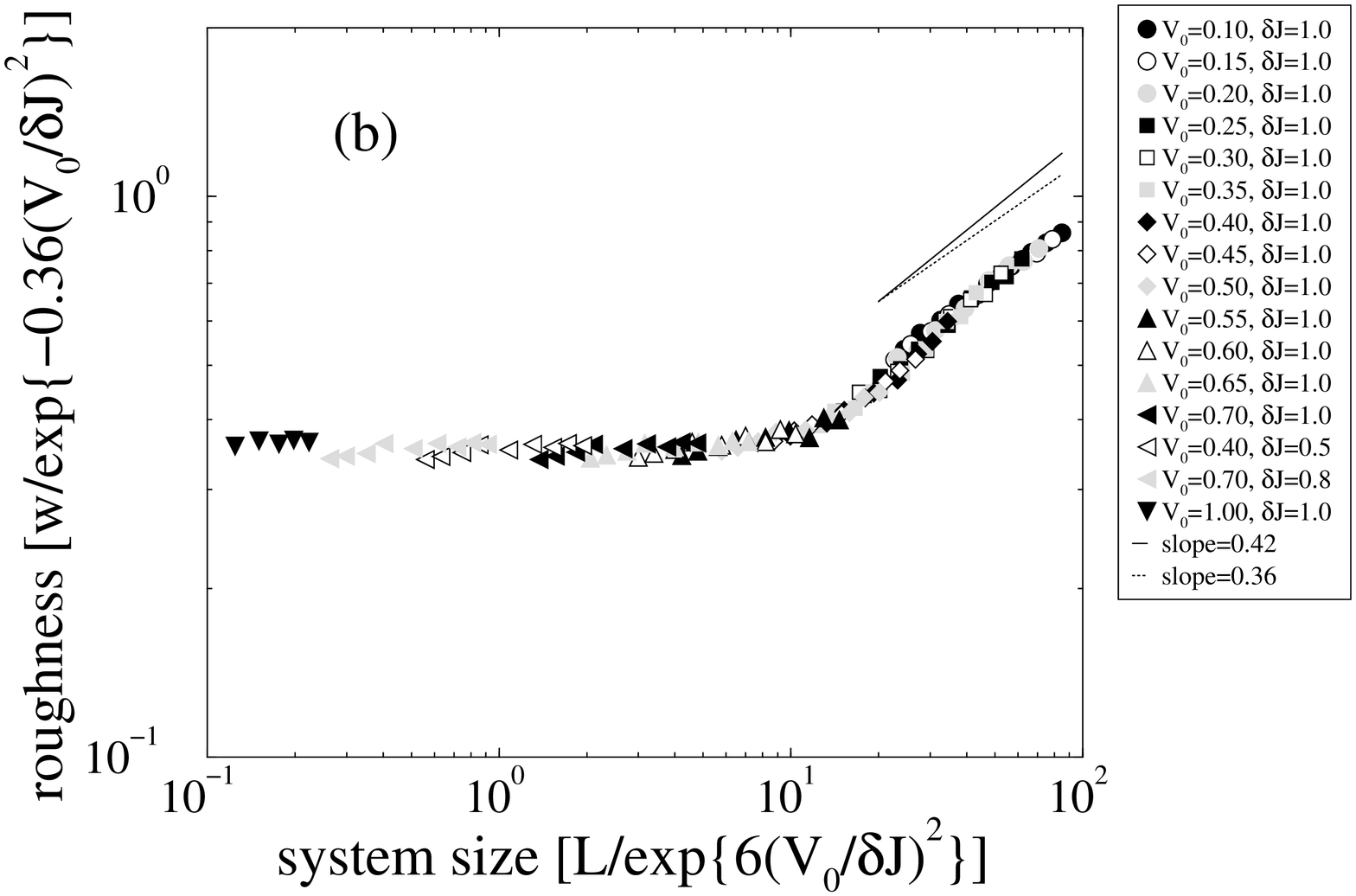,width=7cm,angle=-90}}
\caption{Scaled roughness of interfaces oriented in $\{11\}$ and
$\{111\}$ directions, for various values of $\ V_0\ {\rm and}\ L$.
(a) $\{11\}$ oriented systems with $\lambda=16$, $\delta J=1$, and
system sizes $L^2 = 20^2\--1280^2$.  The number of realizations is 200
for each system size and $V_0$. The solid line corresponds the slope
$\zeta=2/3$. (b) $\{111\}$ oriented systems with $\lambda=4$ and
system sizes $L^3 = 10^3\-- 90^3$.  The number of realizations is 200
for each system size, $\delta J$ and $V_0$. The solid line corresponds
the slope $\zeta=0.42$, while the dotted line is $\zeta = 0.36$. }
\label{fig5}
\end{figure}

The analysis of the last paragraph clearly demonstrates that the
roughening of manifolds in periodic elastic media is via a first order
jump in roughness which is of order the wavelength of the periodic
elastic medium.  It is interesting to investigate whether this first
order jump is observable in the ensemble averaged behavior.  Scaled,
ensemble-averaged plots of the manifold roughness as a function of
$V_0$ are presented in Fig.~\ref{fig5} for \{11\} interfaces
[Fig.~\ref{fig5}(a)] and for \{111\} oriented interfaces
[Fig.~\ref{fig5}(b)].  These plots scale quite nicely with the
characteristic length and roughness suggested by Eqs.~(\ref{Imryjump})
and~(\ref{crossover_l}).  In the two dimensional case, there is also a
clear indication of the first order character of the transition.  The
three dimensional data gives little indication of the first order jump
in roughness and underscores the problems with a naive averaging of
the data. However, we do not have any clear explanation, why the
roughness values in the plateau before the jump can be collapsed with
the same prefactor as in the asymptotic roughness in the \{111\} case,
but not in the \{11\} case.

\section{Periodicity due to the Lattice}\label{100}

\begin{figure}
\centerline{\epsfig{file=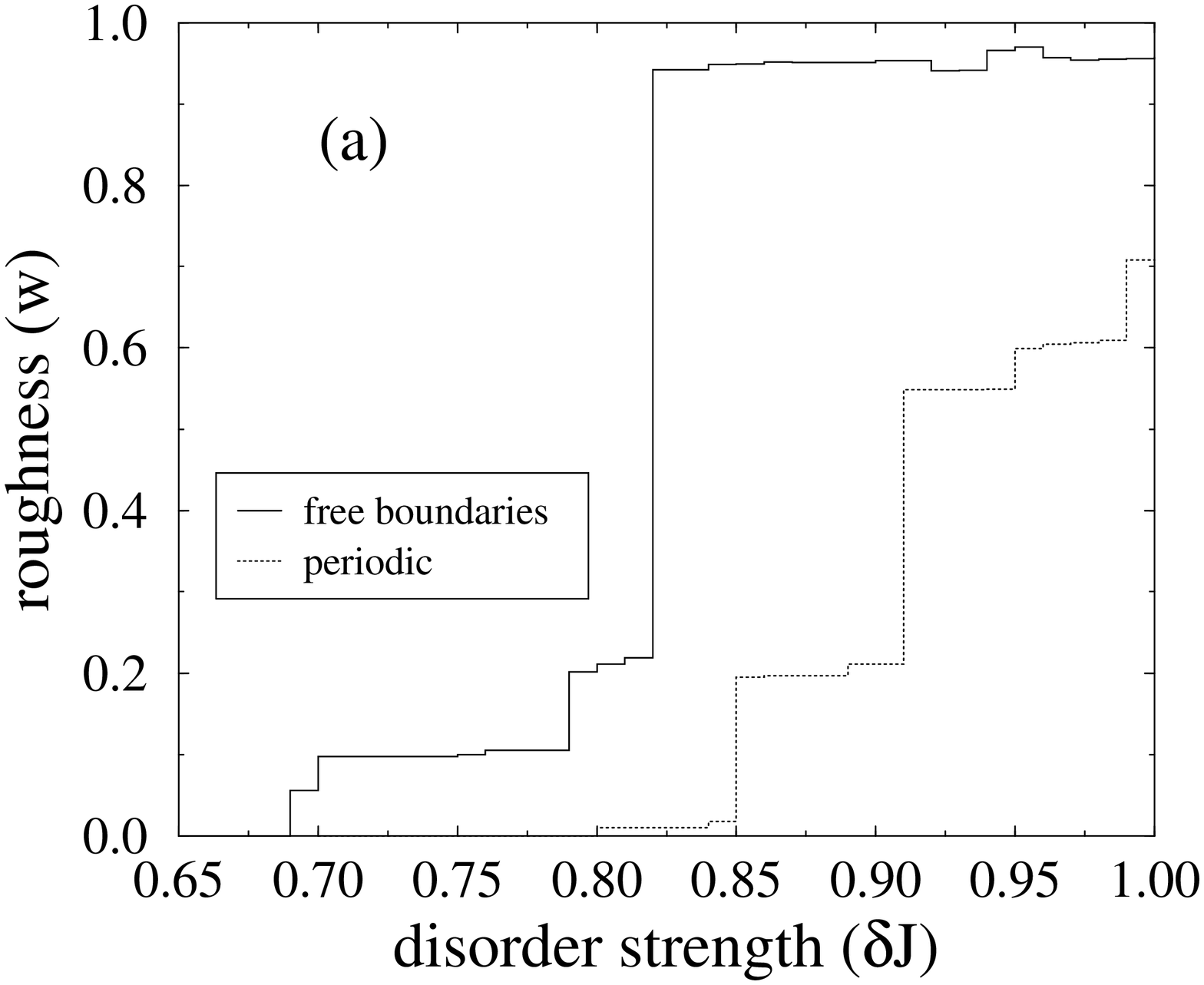,width=7cm,angle=-90}}
\centerline{\epsfig{file=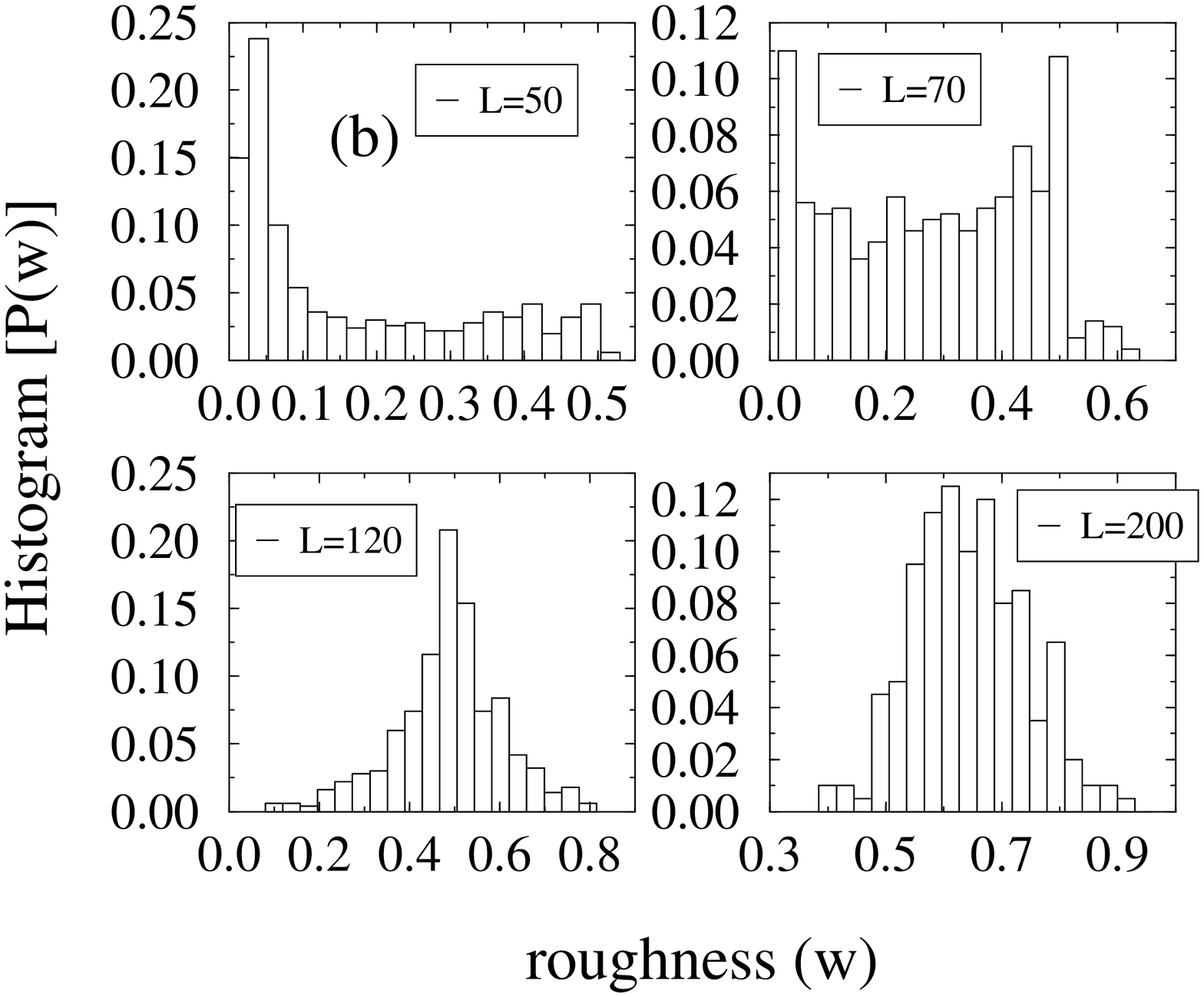,width=7cm,angle=-90}}
\caption{Behavior of the roughness of interfaces oriented in the
$\{100\}$ direction.  (a) The intermittence of a single realization as
a function of the amplitude of uniform disorder $\delta J$. The
disorder configuration is the same (with both free and periodic
boundaries), but the ratio $\delta J/J$ is slowly increased in steps
of 0.01.  The system size is $L^3 = 100^3$. (b) The histograms of the
roughness values $w$ for system sizes $L^2 \times L_h = 50^3 \dots
200^2\times 100 $. The peak of the distributions jumps from $w\simeq
0$ to $w \simeq 0.5$, when the system size increases. The number of
realizations is 500 for smaller system sizes and 200 for $L^2 \times
L_h =200^2\times 100$. $\delta J/J=0.9$.}
\label{fig6}
\end{figure}

\begin{figure}
\centerline{\epsfig{file=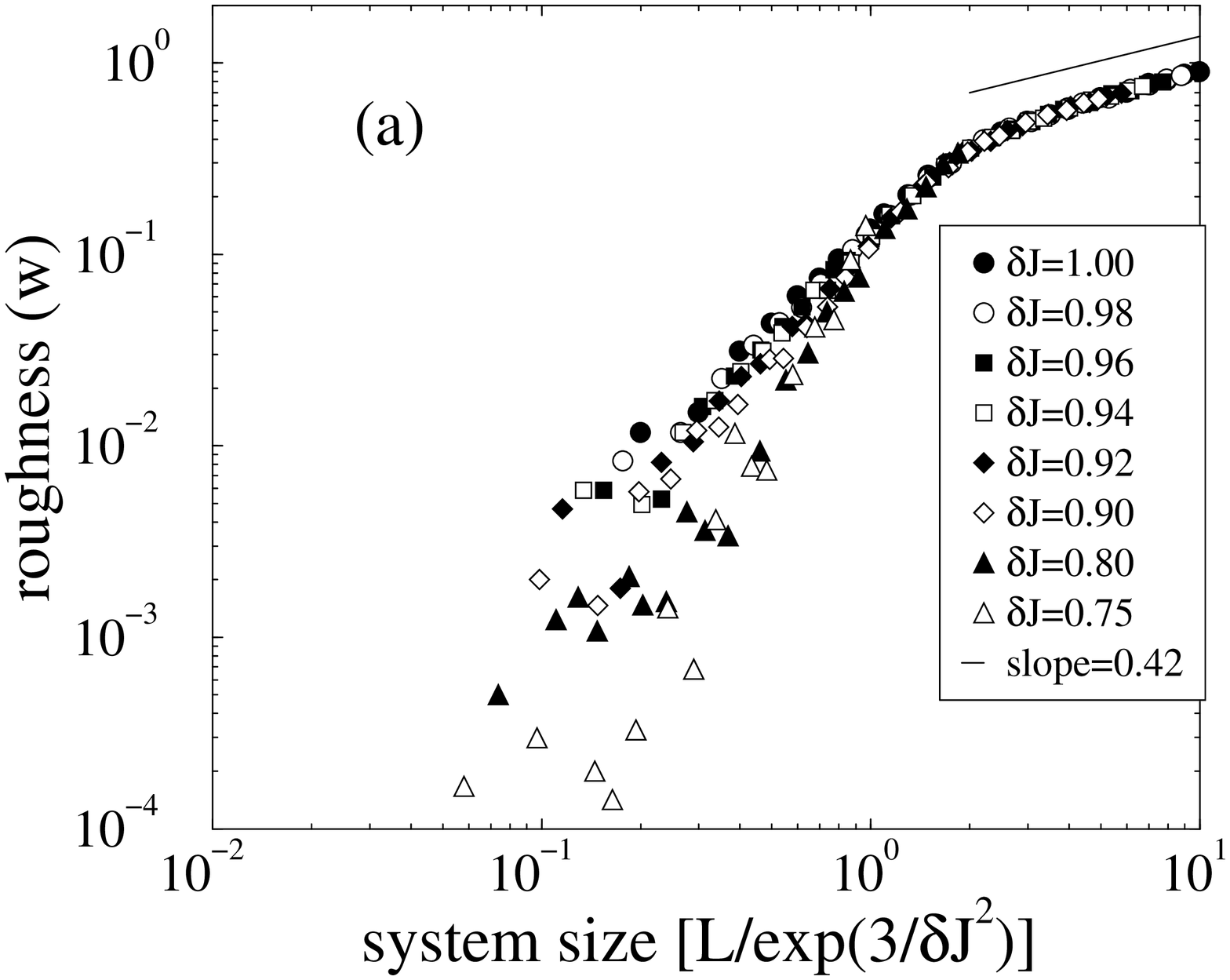,width=7cm,angle=-90}}
\centerline{\epsfig{file=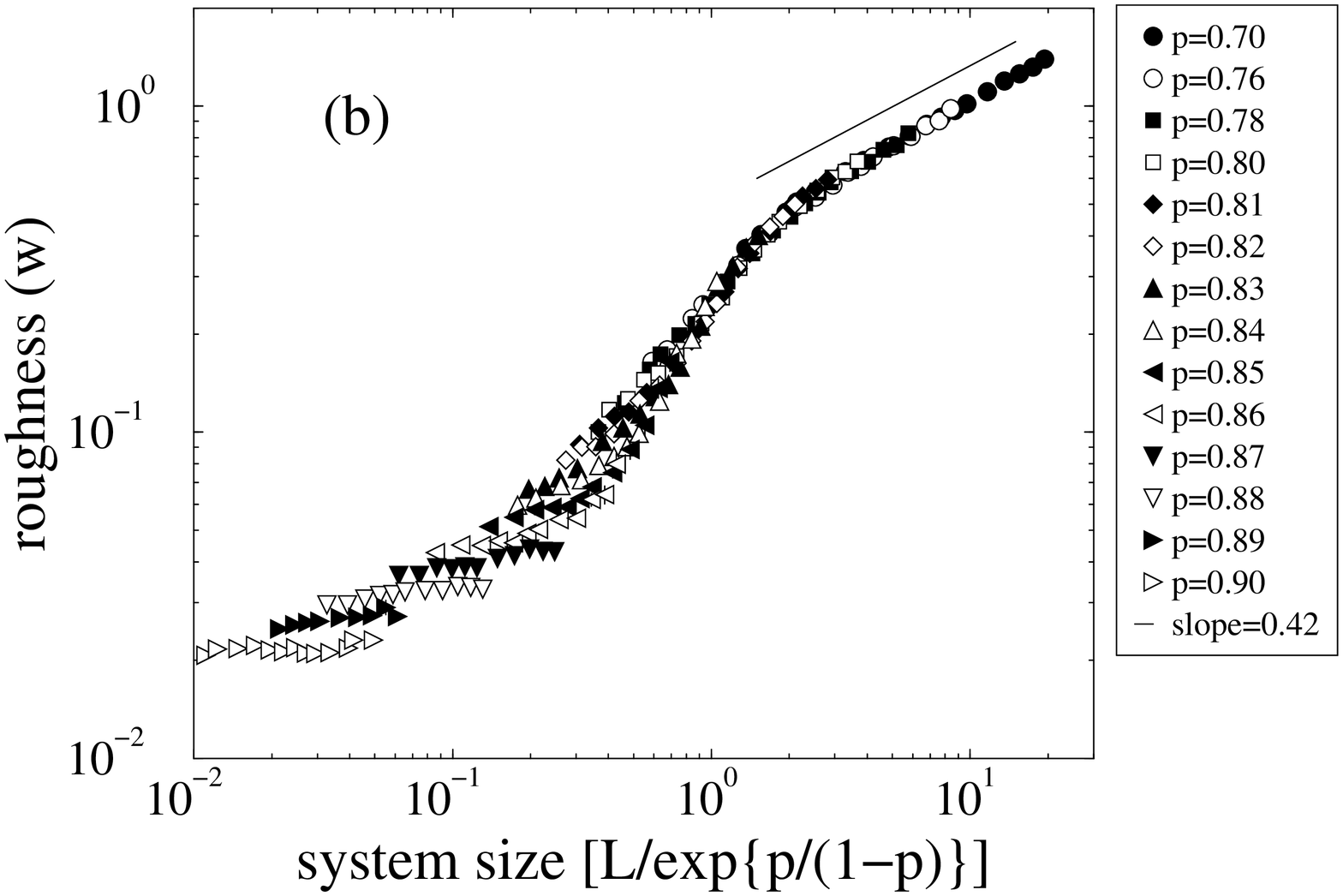,width=7cm,angle=-90}} 
\caption{(a) Scaled roughness for continuum disorder. The system sizes
ranges from $L^2 \times L_h = 4^3$ to $200^2 \times 100$.  The number
of realizations ranges from 500 for system sizes $L^2 \times L_h <
140^2 \times 100$ to 200 for the larger ones. (b) Scaled roughness for
dilution type of disorder. The system sizes ranges from $L^2 \times
L_h = 4^3$ to $200^2 \times 100$ (and even up to $400^3$ for
$p=0.90$). The number of realizations ranges from 500 for system sizes
$L^2 \times L_h < 140^2 \times 100$ to 200 for the larger ones (with
the exception of larger system sizes for $p=0.90$).}
\label{fig7}
\end{figure}

It is of interest to see if the first order character of the
roughening of PEM extends to manifolds in the \{10\} and \{100\}
directions.  In these directions, the lattice itself introduces a
periodicity, which for example is the origin of the thermal roughening
transition in lattice models in three dimensions.  Thus we do not need
to introduce an extra periodic potential, and instead we just study
the roughness of these manifolds as a function of disorder.  We have
studied the roughness of \{100\} manifolds as a function of disorder
before and in those studies we ensemble averaged the data \cite{omat}.
In the light of the understanding develop above, we have revisited
this problem and find that the typical behavior in both the \{10\} and
\{100\} problems is very similar to that suggested by the PEM model.
That is, in a large sample the system roughens via a first order jump
in the roughness due to an extensive fluctuation.  The behavior of one
sample as a function of disorder is presented in Fig.~\ref{fig6}(a).
The probability distributions of the roughness for several $L$
are presented in Fig.~\ref{fig6}(b) in which we observe how one can pass
through a {\it co-existence} region with both flat and rough samples
as $L$ is varied. The intermittent behavior typical of PEM is evident
in Fig.~\ref{fig6}(a), but is obscured by the averaging in
Fig.~\ref{fig7}(a).  Though the jump transition from the flat phase to
the algebraically rough phase occurs in both the periodic elastic
model in the \{111\} direction and for interfaces in the \{100\}
direction, there is an important difference in the behavior of these
models [compare Figs.~\ref{fig5}(b) and~\ref{fig7}(a)].  In the PEM
model in the \{111\} direction, there is a pronounced plateau in the
roughness due to the saturation of wandering within one well
[Fig.~\ref{fig5}(b)].  In contrast, in the \{100\} direction, the
interface remains flat until the transition to the algebraically rough
phase [see Fig.~\ref{fig7}(a)].  The extent of the plateau region can
be tuned in the PEM model by varying the shape of the potential near
the minimum and by varying the wavelength.  We have also carried out
calculations for the case of dilution disorder [Fig.~\ref{fig7}(b)]
and find similar behavior, with the averaged behavior presented in
Fig.~\ref{fig7}(a).  With dilution disorder the pronounced plateau is
not due to any roughening inside a valley, but because of rare
``bumps'', whose occurrence is due to the Poissonian statistics of
diluted bonds.  The averaged data scales quite well with $(\delta
J/J)^2 = p(1-p)J^2/(pJ)^2 = (1-p)/p$, where $J=1$ and the variance of
the binomial distribution $var = std^2 = p(1-p)J^2$ with the
corresponding mean $pJ$, and $p$ is the occupation probability of a
bond.  Thus we find, in contrast to our earlier conclusions from
similar data, that at large enough length scales interfaces in the
\{100\} orientation are algebraically rough and are consistent with
the PEM model.

A further important feature of the large fluctuation character of the
roughening transition is that it is strongly dependent on the boundary
conditions.  This is illustrated in Figs.~\ref{fig2}(b) and
\ref{fig5}(a) in which the roughness is depicted as a function of the
amplitude of the disorder for both periodic and free boundaries and
with the {\it same} disorder configuration.  The threshold value of
$V_0$ at which the first order jump in roughening occurs is typically
smaller for the case of periodic boundaries.  Large fluctuations can
take advantage of the boundary to reduce the cost of crossing the
energy barrier.  This sensitivity to boundary conditions is a hallmark
of the large fluctuation effects discussed here.

\section{Conclusions}\label{concl}

To conclude, we have discussed the roughening of elastic manifolds in
the presence of a competition between bulk randomness and a confining
periodic potential. We have concentrated on the two- and
three-dimensional cases which are well-known to have, asymptotically,
an algebraic roughness scaling. A study of the system-by-system
behavior reveals however a much richer scenario in which each manifold
makes intermittent jumps, finally culminating in a first-order change
in its roughness. This process is also important since it is related
to the asymptotic scaling of the roughness. Recent experiments on the
creep of (1+1)-dimensional systems \cite{Lemerle98} have shown that
scaling arguments of activation energy barriers can match real
systems, using predictions based on rough manifolds.  The time scales
also depend crucially on the actual amplitude which is set in our
picture by the roughening transition.

Also, the intermittence in the early stages would merit experimental
consideration.  Such jumps in the mean location of the interface,
could be studied in the asymptotic rough regime.  In an independent
study we have pointed out this mechanism for both fracture surfaces
arising from random fuse networks, and from yield surfaces of
perfectly plastic media which are equivalent to the minimum energy
surfaces studied here \cite{Sep00b}.

The focus of renormalization group and variational calculations in
this problem has been dimensions $d=(D+1)>4$, since there one
encounters two asymptotic regimes separated by a transition. Of the
two phenomena discussed here at least the intermittent jumps in the
center of mass location of the interface should persist in that case.

\section*{Acknowledgements}\label{acknl}
This work has been supported by the Academy of Finland's Centre of
Excellence Programme (ETS and MJA). PMD thanks the DOE under contract
DOE-FG02-090-ER45418 for support.


\end{multicols}

\begin{references}
\bibitem{HaH95} 
T.\ Halpin-Healy and Y.-C.\ Zhang,
Phys. Rep. {\bf 254}, 215 (1995).

\bibitem{Lemerle98} 
S.\ Lemerle, J.\ Ferr\'e, C.\ Chappert, V.\ Mathet,  T.\ Giamarchi, 
and P. Le Doussal, 
Phys.\ Rev.\ Lett.\ {\bf 80}, 849 (1998); 
P. Chauve, T. Giamarchi, and P. Le Doussal,
Phys.\ Rev.\ B {\bf 62}, 6241 (2000).
T. Nattermann, Y. Shapir, and I. Vilfan, 
Phys.\ Rev.\ B {\bf 42}, 8577 (1990).

\bibitem{Eira2000} 
E.\ T.\ Sepp\"al\"a, M.\ J.\ Alava, and P.\ M.\ Duxbury,
Phys.\ Rev.\ E, {\bf 62}, 3230 (2000).

\bibitem{Bouchaud92} 
J.-Ph.\ Bouchaud and A.\ Georges, 
Phys.\ Rev.\ Lett.\ {\bf 68}, 3908 (1992);
T.\ Emig and T.\ Nattermann
Phys.\ Rev.\ Lett.\ {\bf 81}, 1469 (1998);
T.\ Emig and T.\ Nattermann, 
Eur.\ Phys.\ J.\ B {\bf 8} 525, (1999). 

\bibitem{Fisher86} 
D.\ Fisher, 
Phys.\ Rev.\ Lett.\ {\bf 56}, 1964 (1986);
A.\ A.\ Middleton, 
Phys.\ Rev.\ E {\bf 52}, R3337 (1995).

\bibitem{Parisi90} 
A.\ J.\ Bray and M.\ A.\ Moore,
Phys.\ Rev.\ Lett.\ {\bf 58}, 57 (1987);
G.\ Parisi, 
J.\ Phys.\ France {\bf 51}, 1595 (1990).
M. Mezard, 
J.\ Physique {\bf 51}, 1831 (1990).

\bibitem{Zhang87} 
Y.-C.\ Zhang,
Phys.\ Rev.\ Lett.\ {\bf 59}, 2125 (1987);
T.\ Nattermann, 
Phys.\ Rev.\ Lett.\ {\bf 60}, 2701 (1988);
M.\ V.\ Feigel'man and V.\ M.\ Vinokur,
Phys.\ Rev.\ Lett.\ {\bf 61}, 1139 (1988).

\bibitem{Alavaetal} 
M.\ Alava, P.\ M.\ Duxbury, C.\ Moukarzel, and H.\ Rieger, 
in {\it Phase Transitions and Critical Phenomena}, 
edited by C.\ Domb and J.\ L.\ Lebowitz 
(Academic Press, San Diego, 2001), vol.\ 18.

\bibitem{Goltar88} 
A.\ V.\ Goldberg and R.\ E.\ Tarjan, 
J.\ Assoc.\ Comput.\ Mach. {\bf 35}, 921 (1988).

\bibitem{Galambos} 
J.\ Galambos, {\it The Asymptotic Theory of Extreme Order Statistics},
(John Wiley \& Sons, New York, 1978);
E.\ T.\ Sepp\"al\"a, M.\ J.\ Alava, and P.\ M.\ Duxbury, 
submitted to Phys.\ Rev.\ E.

\bibitem{Seppala00} 
This idea is similar to what happens in the case of random manifolds 
in an external field, see 
E.\ T.\ Sepp\"al\"a and M.\ J.\ Alava, 
Phys.\ Rev.\ Lett.\ {\bf 84}, 3982 (2000).

\bibitem{fisu}
This is analogous to wetting in random systems, see  
R.\ Lipowsky and M.\ E.\ Fisher, 
Phys.\ Rev.\ Lett.\ {\bf 56}, 472 (1986).

\bibitem{Young} 
{\it Spin Glasses and Random Fields}, 
edited by \ A.\ P.\ Young, 
(World Scientific, Singapore, 1997).

\bibitem{Imr75} 
Y.\ Imry and S.-k.\ Ma, 
Phys.\ Rev.\ Lett.\ {\bf 35}, 1399 (1975);
K.\ Binder, 
Z.\ Phys.\ B {\bf 50}, 343 (1983);
also the similar arguments in~\cite{Bouchaud92}.

\bibitem{omat}
M.\ J.\ Alava and P.M. Duxbury,
Phys. Rev. B {\bf 54}, 14990 (1996);
see also 
V.\ I.\ R\"ais\"anen, E.\ T.\ Sepp\"al\"a, M.\ J.\ Alava, and P.\ M.\ Duxbury,
Phys.\ Rev.\ Lett.\ {\bf 80}, 329 (1998).

\bibitem{Sep00b}
E.\ T.\ Sepp\"al\"a, V.\ I.\ R\"ais\"anen, and M.\ J.\ Alava, 
Phys.\ Rev.\ E {\bf 61}, 6312 (2000).
\end{references}
\end{document}